\documentclass[preprint,12pt]{elsarticle}

\usepackage{amssymb}
\usepackage{xcolor}
\usepackage{soul}

\journal{Journal of Theoretical Biology}

\begin{document}

\begin{frontmatter}



\title{Theory of chemical evolution of molecule compositions in the universe, in the Miller-Urey experiment and the mass distribution of interstellar and intergalactic molecules}


\author[label1]{Stuart A. Kauffman}
\author[label2]{Dávid P. Jelenfi}
\author[label2]{Gábor Vattay}
\address[label1]{Institute for Systems Biology, Seattle, WA 98109, USA}
\address[label2]{Department of Physics of Complex Systems, E{\"{o}}tv{\"{o}}s University, 1117 Budapest, P{\'{a}}zm{\'{a}}ny P. s. 1/A, Hungary}



\begin{abstract}
Chemical evolution is essential in understanding the origins of life.  We present a theory for the evolution of molecule masses and show that small molecules grow by random diffusion and large molecules by a preferential attachment process leading eventually to life's molecules. It reproduces correctly the distribution of molecules found via mass spectroscopy for the Murchison meteorite and estimates the start of chemical evolution back to 12.8 billion years following the birth of stars and supernovae. From the Frontier mass between the random and preferential attachment dynamics the birth time of molecule families can be estimated. Amino acids emerge about 165 million years after chemical elements emerge in stars. Using the scaling of reaction rates with the distance of the molecules in space we recover correctly the few days emergence time of amino acids in the Miller-Urey experiment. The distribution of interstellar and extragalactic molecules are both consistent with the evolutionary mass distribution, and their age is estimated to 108 and 65 million years after the start of evolution.  From the model, we can determine the number of different molecule compositions at the time of the emergence of Earth to be 1.6 million and the number of molecule compositions in interstellar space to a mere 719 species.
\end{abstract}




\end{frontmatter}



\section{Introduction}

Reconstruction of the course of chemical evolution can shed light
on fundamental questions of the origins of life. From the order and time in which main types of molecules came into existence, one could infer which molecules coexisted in an evolutionary period. This could be used to estimate how far evolution proceeded from observed molecule associations found in the Universe.
Mass spectroscopy built into recent\cite{quirico2016refractory} and upcoming\cite{moma} space missions
can reveal the chemical makeup of other planets and asteroids.
Electronic and rotation-vibration spectroscopy can identify sets of extragalactic and interstellar molecules\cite{interstellar}, and the Atacama Large Millimeter Array can observe complex organic molecules in other galaxies\cite{sewilo2018detection}. 

Chemical evolution has been studied in the pioneering Miller-Urey, and subsequent experiments\cite{miller1953production,miller1959organic,oro1961amino,parker2011primordial,ferus2017formation}, where amino acids and nucleobases formed from simple, reduced gas mixtures in a matter of weeks. 
Soon after the fall of the Murchison meteorite in 1969, it has been found\cite{wolman1972nonprotein} that non-proteinogenic amino acids produced in the electric discharge experiments were also present in the 4.5 billion years old meteorite, and it contained a racemic mixture\cite{sephton2002organic} pointing to a pre-biotic origin. There are many orders of magnitude difference between the evolutionary timescales in space and the laboratory.   High molecular diversity of the Murchison meteorite has been revealed by mass spectroscopy and where about 58,000 different mass signals have been detected\cite{schmitt2010high}.

Understanding chemical evolution is difficult partly due to the astronomic number of possible molecules and chemical reactions. Describing this vast chemical network seems to be an elusive task. 
The chemical complexity is related to the large number of different molecules which can be built from a given set of chemical elements, the number of different molecules increases super-exponentially\cite{rouvray1996combinatorics,meringer2017exploring} with the size of the set. 

In this paper, we take a new approach and concentrate on the evolution of the masses of molecules only, 
which reduces the complexity of the problem significantly. The mass of
a molecule is the sum $M=\sum_{a=1}^d m_a n_a$, where $m_a$ is the atomic mass of chemical elements, $n_a$ is the number of each element present in the molecule, and $a=1,...,d$ is the index of the chemical element. In this paper we call the vector of integers $\underline{n}=\left[n_1,n_2,...,n_d\right]$ the {\em composition} of the molecule. (It is also known as the {\em molecular formula}, but here it does not have other symbols, such as parentheses, dashes, brackets, commas plus and minus signs, etc.) The combinatorial complexity of possible compositions is much lower than that of the molecules, and it is more readily attainable by experimental methods such as mass spectroscopy.

\section{Results}

\subsection*{Distribution of molecule compositions \label{sec1}} 
The number of linear combinations of atomic masses up to mass $M$ is
\begin{equation}
\sum_{a=1}^d m_a n_a \leq M,
\end{equation}
where $n_a\geq 0$ is an integer. It can be estimated by calculating the
volume
\begin{equation}
\sum_{a=1}^d m_a x_a \leq M,
\end{equation}
where $x_a\geq 0$ is a real number, since a unit volume contains approximately one grid point of the discrete problem. This simplex is a corner of a $d$ dimensional cuboid with side lengths $M/m_a$. The number of possible linear combinations up to a mass $M$ is
\begin{equation}
N(M)\approx \frac{1}{\Gamma(d+1)}(M/M_0)^d\label{eq1}
\end{equation}
where $M_0=(m_1 \cdot m_2\cdot\cdot\cdot m_d)^{1/d}$ is the geometric mean of the masses of the $d$ different atoms and $\Gamma(d+1)=d!$ is the gamma function. It grows sub-exponentially just like a power la. This is an upper bound for the possible number of compositions only, since not all mathematical possibilities do exist as valid molecule compositions. 

\begin{figure}[!ht]
\centering
\includegraphics[width=8cm]{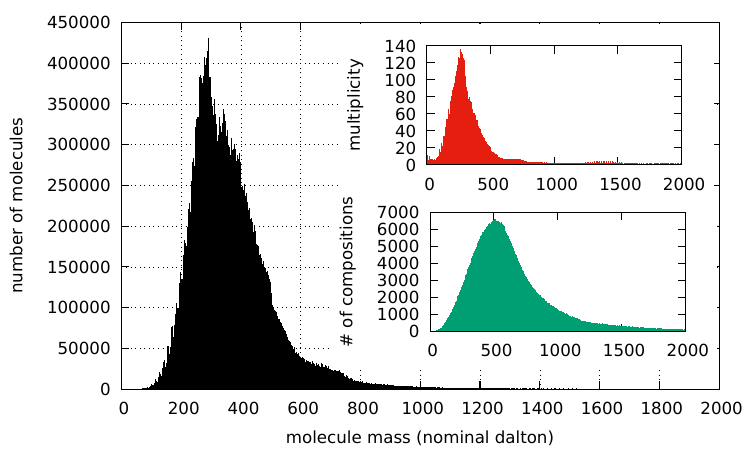}
  \caption{Number of molecule masses in the PubChem compound database (main black), number of compositions  (green inset) and the average number of multiplicity for each composition in a nominal dalton range (red inset). Horizontal axes: masses in nominal dalton (rounded to nearest integer). Vertical axes: counts in each unit dalton range. The molecule mass distribution for small masses increases sharply due to the combinatorial effect as we can see in their average multiplicity.  While the multiplicity increases sharply, the composition distribution increases more moderately. The largest number of molecules is between 290 and 291 Da with 430359 molecules. The largest number of different compositions is between 507 and 508 Da with 6629 compositions. Multiplicity is calculated as the ratio (number of molecules/number of compositions) in a nominal dalton range and peaks between 266 and 267 Da with a multiplicity of 136. For comparison, the number of molecules in the database is about 94 million, and the number of unique compositions is about 3.5 million and $94/3.5\approx 27$ different molecules share the same atomic composition in average.}
  \label{fgr:fig1}
\end{figure}

The PubChem compound database\cite{kim2015pubchem,pubchem}  is 
the largest public databases with about 94 million molecules. While it contains drug molecules synthesized by humans, yet, about $97\%$ of the molecules are unmodified, and can occur in nature. Some compounds having short nucleic acid and amino acid sequences are present, but their share is less than $1\%$. For the molecular mass statistics, we can regard this
dataset as a proxy for all molecules that emerged in pre-biotic chemical evolution existing today. In Fig. \ref{fgr:fig1} (main black) we show the distribution of the molecule masses rounded to the greatest integer in daltons less than or equal to its mass, which we call {\em nominal dalton}. The distribution has a nearly single peak at 290 Da and drops off rapidly in both directions. We can count the number of distinct compositions (i.e., molecules made from the same set of atoms) in each nominal dalton range. We extracted about 3.5 million distinct
molecular compositions, and their distribution is shown in Fig. \ref{fgr:fig1} (green inset). In average 27 molecules are isomers and share the same composition.
This distribution is nearly, and its maximum is less sharp. One can calculate the ratio of the number of molecules and the number of compositions in each nominal dalton range shown in Fig. \ref{fgr:fig1} (red inset). It peaks at 266 Da, where the multiplicity of each composition is 136 on average. The maximum of the composition mass distribution is at 507 daltons. The number of molecules in each nominal dalton range is the product of the number of compositions and their multiplicity. The sharp peak in the number of molecules is
the net result of sharply increasing multiplicity and the more moderately changing number of compositions. 
The multiplicity reflects the combinatorial complexity of molecules, while the composition distribution can be more easily deconstructed on which focus next.

\subsection{The space of compositions}

Chemical evolution starts after the emergence of the chemical elements. Step-by-step, larger and larger molecules can form in the reactions of smaller molecules. In this process, almost all possible small molecules and compositions get created after a certain time. It is reasonable to assume that the number of compositions existing today can be written as a product
\begin{equation}
n(M)=n_0(M)F(M),
\end{equation}
where $M$ is the mass in nominal daltons, $n_0(M)$ is the number of compositions that can exist at a given mass, and $F(M)$ is the fraction of possible compositions that have been created by chemical evolution. The number of possible compositions $n_0(M)$ depends only on the physical quantum chemical properties of molecules and does not depend on the course of chemical evolution.  

\begin{figure}[!htb]
\centering
\includegraphics[width=8cm]{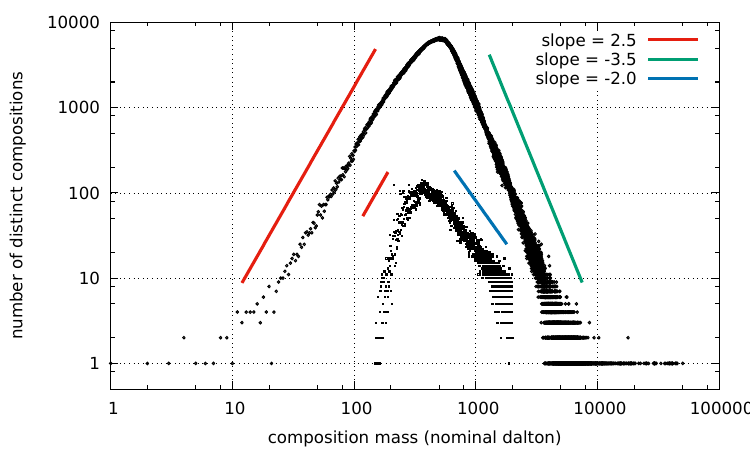}
  \caption{Number of compositions (per nominal dalton) in the PubChem compound database (black circles) and the Murchison meteorite (black squares) on a double logarithmic plot. Both curves start and end with power laws. Slopes of +2.5 (red), -3.5 (green) and -2.0 (blue) are shown for guidance.}
  \label{fgr:fig2}
\end{figure}

For small masses $F(M)\approx 1$, and consequently $n(M)$ coincides with  $n_0(M)$ approximately. From Eq. \ref{eq1} we can expect that 
\begin{equation}
n_0(M) \approx N'(M)=\left(M/M_0\right)^{d-1}/(\Gamma(d)M_0), \label{eq2}
\end{equation}
and grows no faster than a power law. In Fig. \ref{fgr:fig2} (black circles) the composition 
distribution is shown in a double logarithmic plot.  The small mass part starts indeed with a power law with parameters $M_0= 3.81\pm 0.03$ and $d=3.50\pm 0.02$. 
The parameter $d$ can be interpreted as the effective number of elements in compositions. It is consistent with the fact that the molecular space is dominated by organic compounds consisting mainly of three heavy atoms (C, O, N) and the lighter hydrogen, which plays an ancillary role. We develop a theory for the fraction $F(M)$ next.

\subsection{Diffusion in composition space}

The ensemble of molecules can be described by a time dependent probability 
distribution in the composition space $\varrho(\underline{n},t)$, which is governed in general by a nonlinear equation 
$\partial \varrho(\underline{n},t)/\partial t={\cal F}(\varrho(\underline{n},t),\nabla \varrho(\underline{n},t), ...)$,
where ${\cal F}$ is a nonlinear functional of the density and its derivatives. Chemical evolution starts from the elements and the composition space is not populated, $\varrho(\underline{n},0)=0$ for 
compositions involving more than one element. Then, larger and larger molecules can come into existence and the density expands towards larger compositions involving more elements. 
We can assume that the density changes most rapidly at the frontier between the discovered and the yet undiscovered parts of the composition space, where the gradients $\partial\varrho(\underline{n},t)/\partial n_a$ are the largest. If the front expands slowly compared to the time in which chemical reactions reach an equilibrium, the distribution inside of the discovered part of the space is not far from equilibrium and can be described in the framework of a self-consistent approximation. Instead of a detailed description in terms of nonlinear reaction kinetic equations, we can look at the probabilities by which a given molecule composition changes in chemical reactions  
$\underline{n}\rightarrow \underline{n}'$ at equilibrium conditions and can
introduce the equilibrium transition probability $W(\underline{n}\rightarrow \underline{n}')$ and the Master equation
\begin{equation}
\partial_t \varrho(\underline{n},t)=\sum_{\underline{n}'}W(\underline{n}'\rightarrow \underline{n})\varrho(\underline{n}',t) - W(\underline{n}\rightarrow \underline{n}')\varrho(\underline{n},t).
\end{equation}
We can introduce the drifts $\mu_a=\sum_{\underline{n}'}W(\underline{n}\rightarrow \underline{n}')(n_a'-n_a)$ and diffusion coefficients $D_{ab}=\sum_{\underline{n}'}(n_a'-n_a)(n_b'-n_b)W(\underline{n}\rightarrow \underline{n}')$. If the process is dominated by small composition changes of just a few atoms, and the transition probability drops of 
fast for large composition changes, the drifts and the diffusion coefficients are finite, then
then the macroscopic $n_a\gg 1$ behavior of the distribution is governed by the Fokker-Planck equation\cite{van1976expansion}
\begin{equation}
\partial_t \varrho=-\sum_a\partial_ {a}\mu_a\varrho+\sum_{ab}\partial_{a}\partial_{b}D_{ab}\varrho,\label{Fokker}
\end{equation}
where we treat $n_a$ as a continuous variable. 
The total number of each element $N_a=\int n_a\varrho(\underline{n},t)d\underline{n}$ is a conserved quantity in chemical reactions, which can be expressed mathematically as
\begin{equation}
\frac{dN_a}{dt}=\int n_a\partial_t \varrho(\underline{n},t)d\underline{n}=0,
\end{equation}
where the integration goes for the entire composition space. Substituting the time derivative from Eq.\ref{Fokker}, integration by parts, and using the boundary condition of vanishing density $\varrho(\underline{n},)\rightarrow 0$ for sufficiently large compositions $\mid \underline{n}\mid \rightarrow \infty$ leads to the conditions 
\begin{equation}
\int \mu_a(\underline{n})\varrho(\underline{n},t)d\underline{n}=0.
\end{equation}
These should be valid independent of the density yielding $\mu_a(\underline{n})\equiv 0$
and Eq.\ref{Fokker} reduces to the diffusion equation without drifts. The diffusion coefficients depend on the statistics of changes $\Delta n_a=n_a'-n_a$ in chemical reactions. The most probable changes are adding or removing a few atoms in common chemical reaction types, which are the same for small and large molecules, therefore, we can assume that the diffusion coefficients are essentially independent of the composition $\underline{n}$, and are constant practically.
The diffusion constant of the mass is
\begin{equation}
D=\sum_{\underline{n}'}\left(\sum_{a} m_a (n_a'-n_a)\right)^2W(n\rightarrow n')=\sum_{ab}m_a m_b D_{ab}.
\end{equation}
While the number of elements which occur naturally is about $90$, in the previous section we found that the effective dimension of the space of existing compositions is $d=3.5$, and $dN(M)=n_0(M)dM$ is the volume element in this space. 
Assuming homogeneous diffusion in this space, the diffusion equation for the mass is
\begin{equation}
\partial_t\varrho=D\frac{1}{n_0(M)}\partial_M\left(n_0(M)\partial_M \varrho\right),
\end{equation}
where $n_0^{-1}(M)\partial_M n_0(M)$ is the radial part of the divergence operator.
This can be reduced further to $\partial_t\varrho=D M^{-d+1}\partial_M\left(M^{d-1}\partial_M \varrho\right)$.
One can check by direct substitution that 
\begin{equation}
\varrho(M,t)=\frac{1}{(2\pi\sigma^2(t))^{d/2}}\exp\left(-\frac{M^2}{2\sigma^2(t)}\right),
\end{equation}
is a solution of this equation, where $\sigma^2(t)=2Dt$. The probability that a composition with a mass $M$ has been reached by diffusion is 
\begin{equation}
P(m\ge M)=\int_M^\infty  \!\!\!\!\!\!\varrho(m,t)n_0(m)dm\left/ \int_0^\infty \!\!\!\!\!\! \varrho(m,t)n_0(m)dm\right.,
\end{equation}
The average fraction
of the masses discovered by the diffusion process is then
\begin{equation}
F(M)=P(m\ge M)=\frac{\Gamma(d/2,M^2/2\sigma^2(t))}{\Gamma(d/2)},
\end{equation}
where $\Gamma(z,x)=\int_x^\infty \exp(-t)t^{z-1}dt$ is the incomplete gamma function.
Finally, the number of compositions discovered is then
\begin{equation}
n(M)=\frac{1}{\Gamma(d)M_0}\left(\frac{M}{M_0}\right)^{d-1}\cdot\frac{\Gamma(d/2,M^2/2\sigma^2)}{\Gamma(d/2)}.\label{diffusive}
\end{equation}

In Fig. 3 (red) we show that Eq.\ref{diffusive} describes the composition mass distribution extracted from the PubChem database precisely up to about $660$ Da with $\sigma_{PC}=263.3\pm 0.7$ Da.  
Above this critical value this model breaks down, and a power law, and it is described by a different model, which we introduce next. 

\begin{figure}[!hbt]
\centering
\includegraphics[width=8cm]{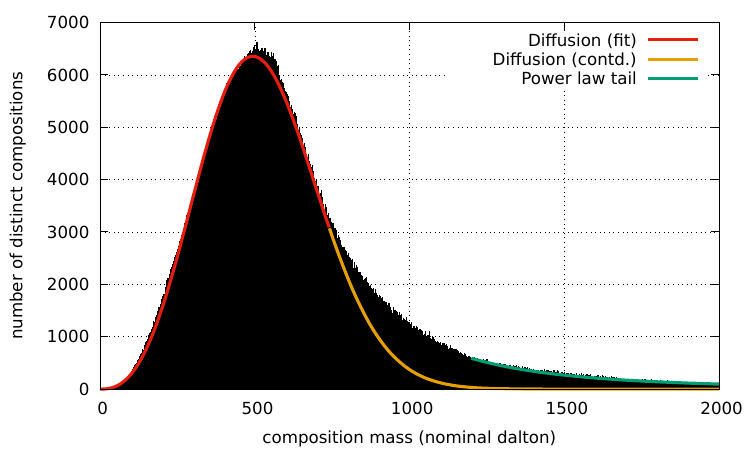}
  \caption{Number of compositions in the PubChem\cite{kim2015pubchem} compound database (black). The main body of the distribution up to 660 Da can be fitted with our diffusive growth based model (red). The rest of the distribution deviates from the model (yellow). The maximum range of the diffusive model is at about 1200 Da, where the power law tail becomes visible (green). In the transitional 660-1200 Da range both diffusion and preferential attachment
  are active, while only preferential attachment can catapult molecules above 1200 Da.}
  \label{fgr:fig3}
\end{figure}

\subsection{Preferential attachment}

So far we have concentrated on the low mass part of the distribution. 
Molecules with some kind of autocatalytic properties grow much faster during the evolution and with various speeds. This is just a few percent in the PubChem database and $n_0(M)$ does not apply to them directly. We assume that the "clock" i.e.: the evolutionary timing can be based on the non-autocatalytic, randomly growing molecules only and we actually used this in the previous
calculations. In this section we concentrate on the "autocatalytic" class of molecules which grow
with some kind of "preferential attachment" i.e. larger molecules are more likely gain mass than small molecules in chemical reactions.

In Fig. 3 (yellow curve) we show the continuation of the diffusion model beyond 660 Da, which drops down rapidly and cannot generate molecules beyond about 1200 Da. In Fig. 2 (black circles) we show that a power law tail is described by $\varrho(M)\sim M^{-3.5}$, where the exponent seems to coincide with the effective dimension $d=3.5$ of the space. This indicates that perhaps a unified model can describe both the low and the high mass parts of the distribution. The high mass tail consists of molecules which have grown faster than in the
simple diffusion process. In the diffusive model, we assumed that molecules grow by a random incremental process. For the large molecules in the tail, this is no longer true. In this region, we can find molecules, which are built from larger building blocks. For example, 
peptide chains or polycyclic aromatic hydrocarbons (PAHs) are not built via random accumulation of atoms, but predominantly from the accumulation of larger blocks such as amino acids and aromatic rings. , and our previous assumptions are not valid here.  As in many growth processes, such as in growing complex networks\cite{barabasi1999emergence,newman2005power} a preferential attachment\cite{simon1955class} process is responsible for the power law tail of the
size distribution. This a 'rich-get-richer' process, in which larger structures can grow
faster than smaller ones. The simplest possible process is when the growth rate is proportional to the size of the existing system, in our case, with the mass of the molecule
\begin{equation}
\frac{dM}{dt}=\lambda M,
\end{equation}
where $\lambda$ is the characteristic rate of the growth process. In such a process the size of the molecule grows exponentially in time, and the random diffusion aggregation process can be neglected. The mass conservation equation in the space of compositions then becomes   
\begin{equation}
\partial_t \varrho+M^{-d+1}\partial_M (M^{d-1}\lambda M\varrho)=0,
\end{equation}
where we used the radial part of the divergence operator $M^{-d+1}\partial_M M^{d-1}$.
The solution of this equation is
\begin{equation}
\varrho(M,t)=e^{-\lambda d t}\varrho_0(Me^{-\lambda t}),
\end{equation}
where $\varrho_0(M)$ is the initial distribution.
It should satisfy the normalization condition
\begin{equation}
\int_1^\infty \varrho_0(M) n_0(M)dM=1,
\end{equation}
where the integration starts at 1 Da to avoid the singularity.
We can have a look at the evolution of an initial power-law $\varrho_0(M)\sim M^\alpha$, which evolves
to $\varrho(M,t)\sim M^\alpha e^{-\lambda (d+\alpha) t}$ since it is an eigenfunction of the problem. 
For $\alpha > -d$ the solution dies out for $t\rightarrow \infty$ and for $\alpha < -d$ the power law 
is not normalizable. The stationary solution is $\alpha = -d$ and 
\begin{equation}
\varrho(M,\infty)=\frac{C}{M^d},
\end{equation}
is normalizable, since
the initial mass distribution vanishes $\varrho_0(M)=0$ above some upper cutoff mass.
Since for large masses the space of compositions is enormous ($\approx 0.5$ million different possible compositions between 1999 and 2000 Da ) it is improbable that the same composition is created via two different reaction pathways, so the number of discovered compositions will be proportional with the density of evolving molecules. Next, we show that not only the power law tail exponent of the distribution can be determined correctly, but the diffusive and the power law regions can be matched precisely.


\subsection{The Frontier of evolution}

In Fig. 3 we can see that the diffusive model and the power law tail join at $M_F\approx 660$ Da, which we call the Frontier. Next, we show that in our model the Frontier is proportional to
the variance and can be estimated as $M_F\approx2.5\sigma$.  
We can get an estimate of from the matching point the diffusive and the power law solutions.
First, we make the crude assumption that the diffusive solution is strictly valid below the Frontier mass and the power law solution is valid above it, and the solutions are continuous
$\varrho_D(M_F)=\varrho_P(M_F)$ and differentiable
$\partial\varrho_D(M_F)/\partial M=\partial\varrho_R(M_F)/\partial M$.
Then, the logarithmic derivatives 
$\partial \ln \varrho(M)/\partial M=(\partial \varrho(M)/\partial M)/\varrho(M)$ should also be equal on both sides.
For the diffusive process the logarithmic derivative of Eq.\ref{diffusive}
\begin{equation}
\frac{\partial\ln\varrho_D(M)}{\partial M}=\frac{d-1}{M}+\frac{\Gamma'(d/2,{M}^2/2
\sigma^2)}{\Gamma(d/2,{M}^2/2\sigma^2)}({M}/\sigma^2).
\end{equation}
For values $M>2\sigma$ the incomplete gamma function can be well approximated with
its asymptotic form $\Gamma(z,x)\approx x^{z-1}\exp(-x)$ and its logarithmic derivative is approximately
$
\partial \ln \Gamma(z,x)/\partial x= \Gamma'(z,x)/\Gamma(z,x)=(z-1)/x -1.
$
Substituting this into the previous expression we get
\begin{equation}
\frac{\partial\ln\varrho_D(M)}{\partial M}=\frac{2d-3}{M}-\frac{M}{\sigma^2}.
\end{equation}
For the power law distribution $\varrho_P(M)=C/M^d$ the logarithmic derivative is
\begin{equation}
\frac{\partial\ln\varrho_P(M)}{\partial M}=-\frac{d}{M}.
\end{equation}
The matching condition 
\begin{equation}
    \frac{2d-3}{M_F}-\frac{M_F}{\sigma^2}=-\frac{d}{M_F},
\end{equation}
then yields $M_F=(3d-3)^{1/2}\sigma=2.74\sigma$, which is not far from the observed numerical
value and confirms the proportionality $M_F\sim\sigma$.

This phenomenological result can be improved from a detailed understanding of the process. In the diffusive region $M<M_F$ a significant fraction of all possible compositions is discovered already. This region is so dense that it is very likely that a new composition created freshly in some reaction will evolve into an already existing composition and becomes indistinguishable from other compositions. Evolution fills in the holes. Therefore, this region is governed by the diffusion model equation irrespective of the individual properties of the molecules. 
We can model the situation by connecting neighboring discovered compositions with links and 
regard this as the network of discovered compositions.
This random network can be characterized by the average degree of the nodes. In $d$-dimension the number of neighbors is $2d$, each of them is already discovered with probability $P(M)=\Gamma(d/2,M^2/2\sigma^2)/\Gamma(d/2)$, and the average degree of the nodes is then
$\langle k \rangle=2dP(M)$. The connectivity of this short range Erd{\H o}s-R\'enyi type network breaks down at the critical point\cite{molloy1995critical}
$\langle k \rangle=\langle k \rangle_c=1$, which is at $P(M_F)=1/2d=1/7$. The solution is
$M_F= 2.48\sigma$ in excellent agreement with our empirical observation. 

This percolation transition\cite{alon1992p} at the border of the densely and the sparsely discovered parts of the composition space is the most exciting area
from an evolutionary perspective. Molecules born freshly in this range in the random growth process can escape the diffusive regime and can enter the preferential attachment regime.

\subsection{The Murchison meteorite}

Once the dimension $d$ and the mass $M_0$ are determined, the number of
possible compositions $n_0(M)$ becomes fully specified, and the diffusive model depends on the variance parameter only, in turn, determined
by the diffusion constant and the time of the diffusion $\sigma^2=2Dt$. To verify this time dependence and the accuracy of the diffusive model, we should compare today's composition distribution to an earlier stage of the evolution. Fortunately, the Murchison meteorite became a reference for extraterrestrial organic chemistry\cite{sephton2002organic}
and about $58,000$ composition has been identified by mass spectrometry\cite{schmitt2010high}. 
\begin{figure}[!htb]
\centering
\includegraphics[width=9cm]{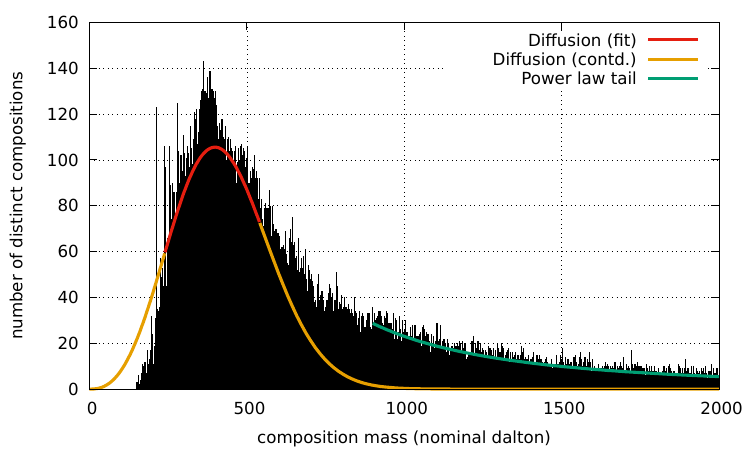}
  \caption{Number of compositions of molecules in the Murchison meteorite identified\cite{reanal} by methanol based electrospray ionization (ESI) mass spectroscopy (black). The composition of masses in the 148-2000 Da range have been measured.
  There is a systematic under-detection below 240 Da since the curve does not reproduce the initial power-law growth of the allowed compositions. In the 240-540 Da range, the diffusive model can be fitted (red). From 540 Da a power law tail develops. The continuation of the diffusive model (yellow) is also shown.}
  \label{fgr:fig4}
\end{figure}
In Fig.\ref{fgr:fig4} we show the number of compositions measured in the Murchison meteorite. Since it is not possible to detect all existing compositions in the sample, we assume that the number of observed mass signals is proportional with the real number
of compositions in the meteorite. The mass range below 148 Da and above 2000 Da is not accessible experimentally. In Fig.\ref{fgr:fig2} (black circles) we show the data in a double logarithmic plot.
We can see that the mass distribution is deviating from the expected power law $M^{2.5}$ for low masses, what we can attribute to the limitations of the experimental technique. The proper trend is recovered around 240 Da, and the central part of the distribution in the 240-540 Da range can be fitted with the diffusive model with $\sigma_{MM}=213.0\pm 1.8 {\rm Da}$. From $M_F=540 {\rm Da}$  a power law tail develops. Note, that the ratio of the observed mass and the fitted variance $M_F/\sigma=2.53$ is in good agreement with our prediction.

The standard estimate\cite{huey1973207pb} for the age of the Murchison meteorite is
$\Delta T_{MM}=T_{EV}-T_{MM}=4.5$ Ga, where $T_{EV}$ the diffusion time from the starting point of the chemical evolution till today and $T_{MM}$ is
the diffusion time till the birth of the Murchison meteorite.
The ratio of the variances is $\sigma^2_{MM}/\sigma^2_{PC}=0.65\pm 0.03$, and
\begin{equation}
T_{EV}=\frac{\Delta T_{MM}}{\left(1-\sigma^2_{MM}/\sigma^2_{PC}\right)},
\end{equation}
yielding $T_{EV}=12.8\pm 0.6$ Ga for the diffusion time from the starting point of the chemical evolution till today. This dates the beginning of chemical evolution to $0.4-1.4$ Ga after the Big Bang in good agreement with the period of nucleosynthesis in stars and supernovae. From the variance and the diffusion time the diffusion constant can also be calculated $D=\sigma^2_{PC}/2T_{EV} = 2708\pm 12 {\rm Da}^2/{\rm Ga}$. 
\begin{figure}
\centering
\includegraphics[width=9cm]{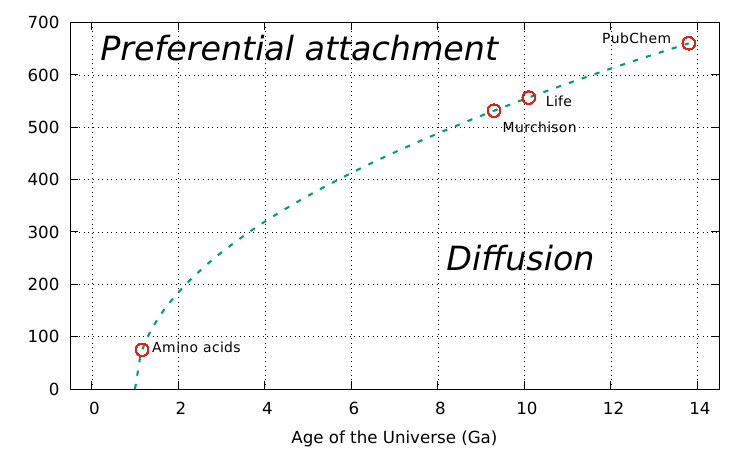}
  \caption{Time-line of chemical evolution. Dashed blue line represents the "Frontier" $M_F=660 (t/T_{EV})^{1/2} {\rm Da}$ between the diffusive and the preferential attachment regimes.  Red dots show the age of amino acids, the Murchison meteorite, the first fossils of life, and the PubChem data today. }
  \label{fgr:fig5}
\end{figure}
In Fig.\ref{fgr:fig5} (dashed blue line) we show the timeline of the evolution of the Frontier of the distribution $M_F=M_F=660 (t/T_{EV})^{1/2} {\rm Da}$. 

Finally, from the Murchison data, we can also get an order of magnitude estimate for the rate $\lambda$. In Fig.\ref{fgr:fig2} the power law tail of the distribution for the Murchison data scales like $ M^{-2}$ and the stationary
tail $M^{-3.5}$ is not even visible even at $M=2000$ Dalton. In the last  $T=4.5$ Ga this initial distribution evolves into the stationary distribution.
We can model the distribution at the beginning with a linear combination of the two
power laws
\begin{equation}
\varrho(M,0)=\frac{C_{3.5}}{M^{3.5}}+\frac{C_{2}}{M^{2}},
\end{equation}
such that we assume that the stationary part is small and not visible $C_2 /(2000)^{2}\gg C_{3.5}/(2000)^{3.5}$. At the end of the evolution the 
distribution becomes
\begin{equation}
\varrho(M,T)=\frac{C_{3.5}}{M^{3.5}}+e^{-\lambda (3.5-2) T}\frac{C_{2}}{M^{2}},
\end{equation}
and in this case the
non-stationary part is non-visible $e^{-\lambda 1.5 T}C_2 /(2000)^{2}\ll C_{3.5}/(2000)^{3.5}$. The two conditions can be written as
$C_2/C_{3.5}\cdot 0.9\cdot 10^5\gg 1$ and $e^{-\lambda 1.5T}C_2/C_{3.5}\cdot 0.9\cdot 10^5\ll 1$.
Assuming that "much greater" means at least a decade in scale in both cases, we get the estimate
\begin{equation}
    e^{-\lambda (1.5) T}\approx 0.01,
\end{equation}
yielding $1/\lambda \approx 1.5$ Ga. 

\subsection{Dating amino acids}

Based on the position of the Frontier we can determine the first appearance of molecule
families. The first member is created at time $T=T_{EV}(M_F/660{\rm Da})^2$, when the Frontier is at the mass of the molecule. Accordingly, glycine ($M=75.0$ Da), the lightest amino acid appears 165 Ma after the start of chemical evolution. The reliability of this result depends on
the long-term accuracy of the diffusion model, i.e., conditions for molecule evolution in the Universe should be relatively stable. The most important factor is the constant supply of atoms from which molecules are formed, especially of carbon. Fortunately, this can be verified
experimentally. The Gemini Near-Infrared Spectrograph (GNIRS) measured the density of intergalactic carbon seeing back to redshifts $z=2-6$ which is approximately the first three billion years of evolution. Stable level of intergalactic carbon on evolutionary timescales has been found\cite{ehrenfreund2011evolution,simcoe2006high} and the density of intergalactic carbon\cite{simcoe2006high} has been determined $\rho_{IG}\approx 3.6 \times 10^{-34} kg/m^3$,
which shows no systematic variation in the entire redshift range. 

To demonstrate the feasibility of the dating of amino acids further, we can compare this result
to experiments in which amino acids spontaneously emerged in the laboratory. In the Miller-Urey experiment, amino acids appear within days. We can compare the timescales of intergalactic chemical reactions to the timescales of these experiments using scaling
relations of reaction rates. In the intergalactic space, the mean free path of reactant particles is
long, the rate-limiting step of the reaction is the deactivation of the collision complex. That is, the reaction is not accomplished until other molecules scatter reactant particles in the vicinity of their partners and lose their excess energy. As the mean free path of reactant particles
becomes longer, they are less frequently scattered in the vicinity of their partners. Accordingly, the rate constant decreases with an increase in the mean free path. It has been shown\cite{tachiya1986influence} that in this case, the reaction rates are proportional to the inverse of the mean free path $k\sim 1/l_{free}$. From this relation, we can get
an order of magnitude estimation for the ratio of reaction rates in the intergalactic space and the Miller-Urey experiment. 
The ratio of reaction rates in the intergalactic space and the Miller-Urey experiments are given by the ratio $k_{MU}/k_{IG}=l_{IG}/l_{MU}$, where $l_{IG}$ and $l_{MU}$ are the mean free paths.
The mean free path is proportional to the average distance of carbon atoms, which in turn is proportional to the third root of the inverse density 
$l_{MU}/l_{IG}=\left(\rho_{IG}/\rho_{MU}\right)^{1/3}$.
In the Miller-Urey experiment, the pressure of the methane gas was $p=2.6 \cdot 10^5 Pa$ corresponding to the carbon density $\rho_{E}= 1.2 kg/m^3$. Using the density of intergalactic carbon we get the ratio $ k_{MU}/k_{IG}\approx 6\cdot 10^{10}$,  meaning that one day of chemical evolution in the Miller-Urey experiment corresponds to $6\cdot 10^{10}$ days evolution in intergalactic space, which is equal to $165$ million years. This almost exact correspondence is of course accidental since
many other factors such as intergalactic radiation may affect the result.

This result indicates that early chemical evolution is closely related to astrochemistry, therefore
we investigate molecules found in the interstellar space next.

\subsection{Interstellar and extragalactic molecules}

\begin{figure}[htb]
\centering
\includegraphics[width=9cm]{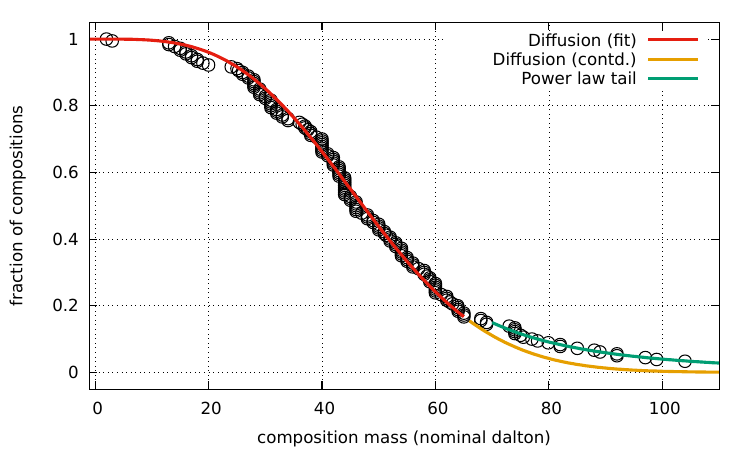}
\caption{Complementary cumulative distribution of 159 compositions that have been detected in the interstellar medium and circumstellar envelopes (black circles). 
Fraction of compositions (vertical axis) whose mass is higher than a given nominal dalton (horizontal axis).
The main body of the distribution up to 51 Da can be fitted with our diffusive growth based model (red). The rest of the distribution deviates from the model (yellow). The range of the diffusive model is at about 90 Da, where the power law tail becomes 
visible (green).}
  \label{fgr:fig6}
\end{figure}
Here we look at the 159 different compositions corresponding to 207 known interstellar and circumstellar molecules\cite{interstellar}. We assume that they represent some early stage of evolution and would like to verify whether the mass distribution follows our distribution.
Due to the small sample size, we analyze the complementary cumulative distribution of the composition masses. The normalized density of composition masses in Eq.\ref{diffusive} is
\begin{equation}
\varrho(M)=\frac{dM^{d-1}}{(2\sigma^2)^{d/2}\Gamma(d)}\Gamma(d/2,M^2/2\sigma^2).
\end{equation}
The complementary cumulative distribution $P_{cc}=\int_M^\infty \varrho(M')dM'$ is
\begin{equation}
P_{cc}(M)=\frac{\Gamma(d,M^2/2\sigma^2)}{\Gamma(d)}-\left[\frac{M^2}{2\sigma^2}\right]^{d/2}\frac{\Gamma(d/2,M^2/2\sigma^2)}{\Gamma(d)}.
\end{equation}
In Fig.\ref{fgr:fig6} we show the complementary cumulative distribution 
of the diffusive model fitted to the distribution up to $M_F=62$ Da with high accuracy. 
The best fit variance is $\sigma_{IS}=24.21\pm 0.05 {\rm Da}$ and the ratio $M_F/\sigma=2.56$ is in excellent agreement with our theory. This dates the sample to $T_{IS}=(\sigma^2_{IS}/\sigma^2_{PC})T_{EV}=108$ Ma after
the start of chemical evolution. Due to the low number of large mass molecules, the tail exponent cannot be determined reliably.  In the tail, we can find large fullerenes, benzene, and benzonitrile. Fullerenes are from the family
of carbon allotropes, which is the first family of breakaway molecules, which grows by a preferential attachment process. Presence of benzene and benzonitrile signals the start of the polycyclic aromatic hydrocarbon (PAH) family and its variants containing eventual heteroatoms. Amino acids are not present, which we can attribute to the fact that this distribution precedes the emergence of the amino acids with about 60 million years.  

\begin{figure}[htb]
\centering
\includegraphics[width=9cm]{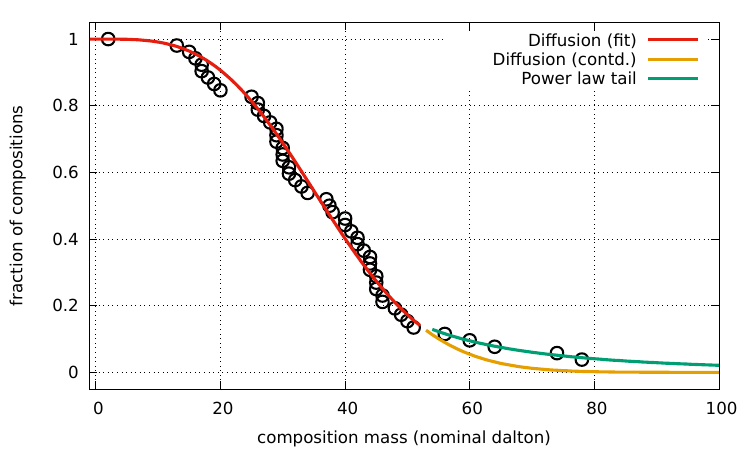}
\caption{Complementary cumulative distribution of 62 extragalactic compositions that have been (black circles). 
Fraction of compositions (vertical axis) whose mass is higher than a given nominal dalton (horizontal axis).
The main body of the distribution up to 51 Da can be fitted with our diffusive growth based model (red). The rest of the distribution deviates from the model (yellow). The range of the diffusive model is at about 70 Da, where the power law tail becomes 
visible (green).}
  \label{fgr:fig7}
\end{figure}
Finally, we analyze known extragalactic molecules\cite{interstellar}. There are 62 different compositions in this set. In Fig.\ref{fgr:fig7} we show the complementary cumulative distribution 
of the diffusive model fitted to the distribution up to $M_F=49$ Da with high accuracy. 
The best fit variance is $\sigma_{XG}=18.78\pm 0.04$ Da and the ratio $M_F/\sigma=2.60$ is in good agreement with our theory, dating the sample to $T_{XG}=(\sigma^2_{XG}/\sigma^2_{PC})T_{EV}=65$ Ma after
the start of chemical evolution. 

The theory makes it also possible to estimate the total number of different compositions existing at a given stage of evolution. Since the diffusive part and the tail match at $M_F=x\sigma$, where $x\approx 2.48$, the number of compositions in the diffusive part and the tail are also linked.
The number of compositions at the Frontier is
\begin{equation}
n(M_F)=\frac{1}{\Gamma(d)M_0}\left(\frac{M_F}{M_0}\right)^{d-1}\cdot\frac{\Gamma(d/2,x^2/2)}{\Gamma(d/2)}.
\end{equation}
From the matching condition the power law tail is
\begin{equation}
n(M)=\frac{M_F^d}{M^d}n(M_F),
\end{equation}
for $M>M_F$. 
The number of compositions in the tail is given by the integral
\begin{equation}
N_T=\int_{M_F}^\infty n(M)dM,
\end{equation}
which yields $N_T=M_F n(M_F)/(d-1).$
The total number of compositions in the diffusive part is the integral
\begin{equation}
N_D=\int_0^{M_F} n(M)dM,
\end{equation}
which gives 
\begin{equation}
N_D=\frac{(x^2/2)^{d/2}\Gamma(d/2,x^2/2)-\Gamma(d,x^2/2)}{M_0^d\Gamma(d/2)\Gamma(d)d}(2\sigma^2)^{d/2}.
\end{equation}
The number of compositions in the diffusive part and in the tail part are then proportional with $\sigma^d$ and
the total number of compositions is then
$N=N_D+N_T\sim \sigma^d\sim t^{d/2}$.
Using the number of compositions today $N=3.5\cdot 10^6$ and the time of evolution
we can fix the constant of proportionality and get for the number of compositions existing at a time
\begin{equation}
N(t)=3.5\cdot 10^6 \left(\frac{t}{T_{EV}}\right)^{1.75}.
\end{equation}
From this, the number of compositions existing at the time of
the Murchison meteorite is about $N_{MM}=1.6\cdot 10^6$. The number of compositions existing in the interstellar space is about $N_{IS}=719$, and the 159 compositions already observed represent about 22$\%$ of them.

\section{Discussion}
We demonstrated that chemical evolution is marching forward robustly in the Universe.
A random diffusive coagulation process produces the main body of the existing molecules while
the tail is the result of a preferential attachment process. The two parts are intimately related
and are separated by the Frontier, where new molecule families are born. Using the PubChem dataset and
the Murchison meteorite mass spectroscopy data we could reconstruct the time evolution and managed to
calculate the time of birth of amino acids, which is about 165 million years after the start of evolution. 
We showed that this is in good agreement with the Miller-Urey experiment after scaling down characteristic times with a factor of $6\cdot 10^{10}$, the ratio of spatial scales. By analyzing the distribution of the composition masses of molecules found in interstellar and extragalactic space, we showed that these molecule assemblies
are described correctly by our distribution, and giving 108 and 65 million years for their time of birth respectively.
We currently don't have an explanation why these distributions froze at those points in time, but the validity of 
the distribution predicts the number of molecules to be found in the interstellar and extragalactic space in the future.
Our findings elevate the role of the statistical analysis of mass spectroscopy signals in future studies of chemical evolution both in laboratory experiments and in the analysis of samples from various parts of the Universe. 
Finally, the results suggest that the main ingredients of life, such as amino acids, nucleotides and other key molecules
came into existence very early, about 8-9 billion years before life. Their existence in samples is by no means
an immediate precursor of life. Life's secrets are coded in the interactions and post-chemical evolution of these molecule families.

After the completion of this manuscript the authors learned about a related paper \cite{wollrab2018miller}, where the evolution of the composition mass density has been measured for the Urey-Miller experiment. Both the shape and the time evolution of the distribution found experimentally is in accordance with the theory outlined here.

\section{Acknowledgements}

The authors thank Istv\'an Csabai, Attila Cs\'asz\'ar, Krist\'of Petrovay, Mark Sephton, Phillippe Schmitt-Koplin, Albrecht Ott and Amri Wandel for illuminating discussions. The autors thank Phillippe Schmitt-Koplin for providing the original data presented in the paper Ref.\cite{schmitt2010high}.
This research was supported by the National Research Development and Innovation Office of Hungary (Project No. 2017-1.2.1-NKP-2017-00001) and the ELTE Excellence Program (783-3/2018/FEKUTSRAT). G.V. is funded partially by Novo Nordisk Foundation (16584).



\bibliographystyle{elsarticle-harv}


\end{document}